\documentclass[a4paper,fleqn,usenatbib]{mnras}

\usepackage{newtxtext,newtxmath}

\usepackage[T1]{fontenc}
\usepackage{ae,aecompl}

\usepackage{graphicx}	
\usepackage{amsmath}	
\usepackage{amssymb}	
\usepackage{hyperref}

\usepackage{adjustbox}
\usepackage{ulem}

\def\be{\begin{equation}}
\def\ee{\end{equation}}
\def\ba{\begin{eqnarray}}
\def\ea{\end{eqnarray}}
\def\m{\mathrm}

\def\Mdotin{\dot{M}_{\mathrm{in}}}

\def\Mdot{\dot{M}}

\def\Pdot{\dot{P}}

\def\Msun{M_{\odot}}

\def\Lcool{L_{\mathrm{cool}}}
\def\Lacc{L_\mathrm{acc}}

\def\rin{r_{\mathrm{in}}}
\def\rlc{r_{\mathrm{LC}}}
\def\rout{r_{\mathrm{out}}}
\def\rco{r_{\mathrm{co}}}

\def\Lx{L_{\mathrm{x}}}

\def\Md{M_{\mathrm{d}}}

\def\r_m{r_\mathrm{m}}
\def\v_esc{v_\mathrm{esc}}

\def\rA{r_{\mathrm{A}}}

\def\Tp{T_{\mathrm{p}}}

\def\Teff{T_{\mathrm{eff}}}

\def\Firr{F_{\mathrm{irr}}}

\def\P0min{P_{0,{\mathrm{min}}}}

\def\tc{\tau_{\mathrm{c}}}

\def\m{\mathrm}

\def\cs{c_{\mathrm{s}}}

\def\Alfven{Alfv$\acute{\mathrm{e}}$n~}

\def\ergpers{erg~s$^{-1}$}

\def\spers{s~s$^{-1}$}

\def\B0-P{$B_0$~--~$P$}
\def\PPdot{$P$~--~$\Pdot$}

\title[On the long-term evolution of RRATs]{On the long-term evolution of rotating radio transients}

\author[A. A. Gen\c{c}ali et al.]{
A. A. Gen\c{c}ali,$^{1}$\thanks{E-mail: gencali@sabanciuniv.edu}
\"{U}. Ertan,$^{1}$
\\
$^{1}$Sabanc{\i} University, Orhanl{\i} Tuzla, 34956 \.{I}stanbul, Turkey
}

\date{Accepted XXX. Received YYY; in original form ZZZ}

\pubyear{2020}

\begin{document}
\label{firstpage}
\pagerange{\pageref{firstpage}--\pageref{lastpage}}
\maketitle

\begin{abstract}
Investigation of the long-term evolution of rotating radio transients (RRATs) is important to understand the evolutionary connections between the isolated neutron star populations in a single picture. The X-ray luminosities of RRATs (except one source) are not known. In the fallback disc model, we have developed a method to estimate the dipole field strengths of RRATs without X-ray information. We have found that RRATs could have dipole field strengths, $B_0$, at the poles ranging from $\sim 7 \times 10^9$~G to $\sim 6 \times 10^{11}$~G which fill the gap between the $B_0$ ranges of central compact objects (CCOs) and dim isolated neutron stars (XDINs) estimated in the same model. In our model, most of RRATs are evolving at ages $(\sim 2$~--~$6) \times 10^5$~yr, much smaller than their characteristic ages, such that,  cooling luminosities of a large fraction of relatively nearby RRATs could be detected  by the eROSITA all-sky survey. Many RRATs are located above the upper border of the pulsar death valley with the fields inferred from the dipole-torque formula, while they do not show strong, continuous radio pulses. The $B_0$ values estimated in our model, place all RRATs either into the death valley or below the death line. We have tentatively proposed that RRATs could be the sources below their individual death points, and their short radio bursts could be ignited by the disc-field interaction occasionally enhancing the flux of open field lines through the magnetic poles. We have also discussed the evolutionary links between CCOs, RRATs and XDINs.
\end{abstract}




\section{Introduction}
\label{sec:intro}

	Rotating radio transients (RRATs) were discovered more than a decade ago through single-pulse search reprocessing of the Parkes Multi-beam Pulsar Survey data \citep{Mc2006}. Unlike ordinary radio pulsars, RRATs show sporadic and short radio bursts with time separations of $\sim$ minutes to a few hours. Durations of the radio bursts vary from $0.5$~ms to $100$~ms with very high flux densities in the $10^{-3}$~--~$10$~Jy range. It was proposed by \citet{Weltevrede2006} that these sources could be emitting non-detected continuous weak radio pulses, in addition to their sporadic radio bursts. Recently, weak radio pulsations were indeed observed from several RRATs \citep{Cui2017, Bhattacharyya2018}. Out of more than $100$ known RRATs, period derivatives were measured for 34 sources (see Table~\ref{tab:PPdotLx}). Only one of these systems (PSR J1819--1458) was detected in X-rays \citep{Rea2009}, and there are X-ray upper limits for 5 sources (Table~\ref{tab:PPdotLx}). There are uncertainties in the positions and distances of RRATs that prevent detections of these systems in X-rays \citep{Kaplan2009}. Some of RRATs could have very low X-ray luminosities that remain below the detection limits. Their spin periods and period derivatives are in the ranges of $0.1$~s~$ < P < 8$~s and $10^{-16}$~\spers~$ < \Pdot < 10^{-12}$~\spers~respectively, which give characteristic ages, $\tc \sim P/ 2 \Pdot$, under the usual scenario of magnetic dipole braking in vacuum varying from $10^5$~yr to $10^8$~yr.

	Physical conditions producing the radio bursts of RRATs is not well known. Several ideas were proposed to explain the RRAT properties: The mechanism of these bursts might be related to those running in the systems that show giant radio pulses \citep{Knight2006}, or nulling pulses with extreme nulling fractions \citep{Redman2009}. Presence of circumstellar material or radiation belts affecting the charged density in the magnetospheres of neutron stars could produce the radio bursts  \citep{Cordes2008}. Alternatively, this behavior might be the characteristics of the neutron stars that are close to the pulsar death line in the magnetic dipole field--period (\B0-P) plane \citep{Bhattacharya1992, Chen1993, Zhang2007}. We use “$B_0$” to denote the dipole field strength at the pole of the neutron star.

	Depending on the dipole field strength and geometry, continuous radio pulsations of a pulsar are expected to terminate at a point inside a death valley in the \B0-P diagram \citep{Chen1993}. The lower boundary of the death-valley is similar to the classical pulsar death line \citep[see][and references therein]{Bhattacharya1992} below which pulsed radio emission is not expected. Some of the radio pulsars could die close to the upper boundary, while some others switch off the radio pulses close to the death line (lower boundary). 
	
 	If the external torque dominates the dipole torque, the constant-$B_0$ lines in the \PPdot~diagram overestimate the actual $B_0$. In this case, for instance, the $B_0$ value inferred from the dipole torque formula could place the source inside or above the upper boundary of the death valley, while the source could actually be below the death line. In the magnetar model \citep{Thompson1995}, like for other isolated neutron star populations, RRATs are assumed to slow down by dipole torques alone. In other words, in this model, most of these sources are expected to be powerful and continuous pulsed radio emitters. In this picture, the possibility that RRATs could be the sources close to death line, or to the death “points” inside the death valley is excluded from the alternative explanations for the RRAT behavior.  
	
	Fallback matter from supernova explosions is estimated to form discs around the newly formed compact objects \citep{Colgate1971, Michel1988, Chevalier1989}. From the numerical simulations, \citet{PERNA2014} estimated that fallback discs around neutron stars are not likely outcome of a supernova, unless the explosion is weak or highly anisotropic. It was proposed that the X-ray luminosities and rotational properties of anomalous X-ray pulsars (AXPs) could be accounted for by evolution of young neutron stars with fallback discs \citep{Chatterjee2000}. \citet{Alpar2001} proposed that the properties of fallback discs, if added to the initial conditions together with magnetic dipole moment and initial period, could explain the emergence of diverse neutron star populations. The fallback disc model was used to explain different rotational properties of isolated neutron stars that are not explained by evolutions with dipole torques alone \citep{Marsden2001, Menou2001, Eksi2003, Yan2012, FuLi2013}. The model proposed by \citet{Alpar2001} was developed later considering the effects of the X-ray heating and inactivation of the disc on the long-term evolution \citep{ErtanE2009}. From the earlier applications of this model to different neutron star populations \citep[see e.g.][]{Ertan2014}, we estimate that there could be evolutionary links between different systems, which is also important in estimating the birth-rate of isolated neutron stars \citep{Popov2006, KeaneK2008}. 
	
	Emission characteristics of the fallback discs were also studied comprehensively \citep{Perna2000, Ertan2006, Ertan2007, Ertan2017, Posselt2018}. The broad-band spectrum of 4U 0142+61 from the optical to mid-IR bands \citep{Hulleman2000, Hulleman2004, Morii2005, Wang2006} is in good agreement with the emission from the entire disc surface \citep{Ertan2007}. Mid-IR emission at $4.5~\mu$m was detected from 1E 2259+58 as well which is similar to that of 4U 0142+61 indicating an emission mechanism similar to that of 4U 0142+61 \citep{Kaplan2009ApJ}.
	 		
	The rotational properties of a neutron star evolving with a fallback disc \citep{Chatterjee2000, Alpar2001, ErtanE2009} are governed mainly by the magnetic torque arising from the interaction between the inner disc and the dipole field of the star. In this model, magnetic dipole moments estimated for isolated neutron star populations are one to three orders of magnitude smaller than those inferred from the dipole torque formula. For instance, the $B_0$ ranges that can produce the rotational properties and the X-ray luminosities of the sources are ($\sim 10^{12}$~--~$10^{13}$)~G for anomalous X-ray pulsars (AXPs) and soft gamma repeaters (SGRs) \citep{Benli2016}, and ($\sim 10^{11}$~--~$10^{12}$)~G for dim isolated neutron stars (XDINs) \citep{Ertan2014}, while the dipole torque formula gives $B_0 > 10^{13}$~G for all XDINs, and $B_0 > 10^{14}$~G for most of AXP/SGRs. It is important to note here that the energetic bursts of SGRs require magnetar fields as described in the magnetar model. Presence of these strong fields are compatible with the disc model provided that they are in quadrupole components which have small scales and are located close to the neutron star \citep[see e.g.][]{Alpar2011}. Since the disc interacts with the large-scale dipole component of the field, the higher multipoles do not affect the rotational evolution of the star. Discovery of low-B magnetars \citep{Rea2010, Livingstone2011, Rea2012} clearly showed that the SGR bursts do not require magnetar dipole fields.
		
	In the fallback disc model, earlier work on the evolution of neutron star populations indicates that most of the AXP/SGRs are in the accretion with spin-down (ASD) phase at present, while XDINs were spun down to long periods in the ASD phase, and are currently evolving in the strong-propeller (SP) phase. In the SP phase, there is no accretion on to the star, and the sources are allowed to emit radio pulses. Nevertheless, the $B_0$ values estimated in the model together with the observed periods place all known XDINs well below the pulsar death line \citep[see][ fig.~4]{Ertan2014}. The lack of pulsed radio emission, which could be attributed to narrow beaming and the viewing geometry in the dipole torque model, is due to insufficient field strengths of XDINs in the fallback disc model.
 	
	To investigate the details of the long-term evolution and the current ages of RRATs is not easy, since their X-ray luminosities are not known. Recently, we studied the evolution of PSR J1819--1458 (hereafter J1819), the only RRAT that was detected in X-rays, in the fallback disc model. Our results showed that the rotational properties and $\Lx$ of J1819 can be achieved simultaneously by a neutron star evolving with a fallback disc \citep{Gencali2018}. Simulations indicate that J1819 is currently in the ASD phase, close to the ASD/SP transition, and the source is evolving toward the XDIN properties which might imply an evolutionary link between RRATs and XDINs \citep{Gencali2018}. The source properties can be reproduced with $B_0 \sim 5 \times 10^{11}$~G which places J1819 below the pulsar death line. This means that J1819 is not likely to emit regular radio pulses even after its accretion phase terminates. The other RRATs are distributed over large ranges of $P~(0.1$~--~$8$~s) and $\Pdot~(10^{-16}$~--~$10^{-12}$~\spers) with unknown X-ray luminosities.

	In our long-term evolution model, can we estimate the dipole moments of RRATs without any $\Lx$ information? This is possible, if we somehow determine the evolutionary phases (ASD or SP) of the sources. In the earlier work, the $B_0$ values estimated for AXP/SGRs, XDINs, high-B radio pulsars (HBRPs), and central compact objects (CCOs) \citep{Ertan2014, Benli2016, Benli2017, Benli2018, BenliCCO2018} in the same model range from a few $10^9$~G (for CCOs) to several $10^{12}$~G (for AXP/SGRs and HBRPs). There is a $B_0$ gap indicated by these results between the maximum $B_0$ of CCOs ($\sim 4 \times 10^9$~G) and the minimum $B_0$ of XDINs ($ \sim 3 \times 10^{11}$~G). In this work, we will investigate whether the $B_0$ range of RRATs could fill this gap in the fallback disc model. We briefly describe the model in Section~\ref{sec:model}. The details of the field estimation for RRATs are given in Section~\ref{sec:evol}. We discuss our results in Section~\ref{sec:conc}, and summarize our conclusions in Section~\ref{sec:SumofResults}.

\section{The model}
\label{sec:model}

	We use the long-term evolution model for a neutron star with a fallback disc which is applied earlier to other young neutron star populations \citep{ErtanE2009, Caliskan2013, Ertan2014, Benli2016, Benli2017, Benli2018, BenliCCO2018, Gencali2018}. Since the details of the model are given in these work, we briefly describe the model calculations below.
		
	We solve the disc diffusion equation using the $\alpha$-prescription for the kinematic viscosity, $\nu = \alpha \cs h$, where $\alpha$ is the kinematic viscosity parameter, $\cs$ is the sound speed, and $h$ is the pressure scale-height of the disc \citep{Shakura1973}. We start with a surface density profile for a steady, geometrically thin disc. In the accretion with spin-down (ASD) phase, the matter coupling to the field lines from the co-rotation radius, $\rco$, at which the speed of the field lines equals the Kepler speed, flows along the field lines on to the star. In this phase, we equate the accretion rate, $\Mdot_\ast$, to the mass-inflow rate of the disc, $\Mdotin$, which gives an X-ray luminosity $\Lacc = G M \Mdot_\ast/R_\ast$ where $G$ is the gravitational constant, $M$ and $R_\ast$ are the mass and the radius of the neutron star. We calculated the total X-ray luminosity, $\Lx$, including the contribution of the intrinsic cooling luminosity, $\Lcool$, of the star. We use the theoretical $\Lcool$ curve for neutron stars with conventional magnetic dipole fields \citep{Page2009}. In the calculation of $\Lcool$, we also include the internal heating by external torques \citep{Alpar2007} which does not affect significantly the $\Lx$ evolution of young neutron stars with fallback discs. When the accretion is allowed, $\Lacc$ dominates $\Lcool$ in most cases.
		
	 In addition to viscous dissipation, the disc is heated by the X-rays emitted by the star. Effective temperature of the disc can be written as $\Teff \simeq \Big[ (D + \Firr)/\sigma \Big]^{1/4} $ where $\sigma$ is the Stefan-Boltzmann constant, and D is the rate of viscous dissipation per unit area of the disc. The irradiation flux can be written as $\Firr = 1.2~C \Lx / ( \pi r^2 )$ where $r$ is the radial distance from the star, and $C$ is the irradiation parameter which depends on the albedo and geometry of the disc surfaces \citep{Fukue1992}. Outside a radius of about $10^9$~cm, $\Firr$ dominates $D$. There is a critical temperature, $\Tp$, below which the disc becomes viscously inactive. Dynamical outer radius of the active disc corresponds to the radius at which $\Teff$ currently equals $\Tp$, that is, $\rout = r(\Teff = \Tp)$. In the long-term evolution, $\rout$ decreases gradually with slowly decreasing $\Firr$. In the earlier work on different neutron star populations, the simulations with $\Tp \sim 100$~K, and $C$ in the $(1$~--~$7) \times 10^{-4}$ range reproduced the individual source properties \citep{Ertan2006, Ertan2007, Caliskan2013, Ertan2014, Benli2016, Benli2017, Benli2018, BenliCCO2018, Gencali2018, Ozcan2020}. The $\Tp$ values estimated in the model seem to be consistent with the results of the theoretical work indicating that the disc is likely to be active at very low temperatures \citep{Inutsuka2005}. The $C$ values are in the same range as that estimated for the low-mass X-ray binaries \citep[see e.g.][]{Dubus1999}. 
					
	In the ASD regime, we calculate the spin-down torque, $N_\m{D}$, produced by the disc-field interaction, by integrating the magnetic torques between the co-rotation radius, $\rco = (GM/ \Omega_\ast^2)^{1/3}$, and the conventional \Alfven radius, $\rA \simeq \Big[ \mu^4 / (GM \Mdotin^2) \Big] ^{1/7}$, where $\mu$ and $\Omega_\ast$ are the magnetic dipole moment and angular frequency of the neutron star. This allows us to write $N_\m{D}$ in terms of $\Mdotin = \Mdot_\ast$ and $\rA$ as $N_\m{D} = \frac{1}{2} \Mdotin (G M \rA)^{1/2} [1-(\rA/\rco)^3 ]$ \citep{Ertan2008}. In this phase, $\rco < \rA < \rlc$, where $\rlc = c/\Omega_\ast$ is the light cylinder radius, and $c$ is the speed of light. The accretion on to the star from $\rco$ produces a spin-up torque, $N_\m{acc} = \Mdot_\ast (GM\rco)^{1/2}$. The total torque can be written as $N_\m{TOT} = N_\m{D} + N_\m{acc} +N_\m{dip} $, where $N_\m{dip} = -2 \mu^2 \Omega_\ast^3 / 3c^3$ is the magnetic dipole torque. In this phase, contributions of $N_\m{acc}$ and $N_\m{dip}$ are usually negligible compared to $N_\m{D}$ for the accretion regimes of AXP/SGRs, XDINs, CCOs, and HBRPs except few sources that are very close to rotational equilibrium with relatively high $\Lx$ \citep{Ertan2014, Benli2016, Benli2017, Benli2018, BenliCCO2018}.
	
	 In the strong-propeller (SP) phase, assuming that all the inflowing matter is thrown out of the system from the inner disc, we take $\Mdot_\ast = 0$, $N_\m{acc} = 0$, and $\Lx = \Lcool$. In the absence of a well known critical condition for the transition to the SP phase at low $\Mdotin$, we assume that this transition takes place when $\rA = \rlc$, and the system is in the SP phase for lower $\Mdotin$ that gives $\rA > \rlc$. Since $\rA$ greater than $\rlc$ has no physical meaning, we replace $\rA$ in the $N_\m{D}$ equation with $\rlc$ in the SP phase. The $\Mdotin$ curve enters a sharp decay phase at the end of the ASD regime, and the accretion-propeller transition is likely to take place in this sharp decay phase. This means that the exact value of the critical $\Mdotin$ for this transition does not affect our results significantly. A more realistic model to estimate the ASD/SP transition condition was developed by \citet{Ertan2017}. In this model, the critical $\Mdotin$ values for this transition are close to those obtained from our simplified model. We prefer to use the same simplified model for RRATs as well for a systematic comparison of initial conditions of these sources with those obtained earlier for the other isolated neutron star systems \citep{Caliskan2013, Ertan2014, Benli2016, Benli2017, Benli2018, BenliCCO2018, Gencali2018}. This comparison also requires to use a similar set of main disc parameters, $\alpha$, $\Tp$, and $C$ for different populations, since these parameters are expected to be similar for the fallback discs of different neutron star populations. In this way, we are able to compare the evolutionary paths leading to very different source properties resulting only from the changes in the initial conditions, namely the initial period, $P_0$, the magnetic dipole strength on the pole of the neutron star, $B_0$, and the initial mass of the disc, $\Md$. 
	 
	 Illustrative model curves showing the effects of the initial conditions on the evolutionary curves are plotted in the earlier works \citep{ErtanE2009, Ertan2014, Benli2015}. We can summarize the effects of the initial conditions ($P_0$, $B_0$, and $\Md$) as follows: For the sources that enter the evolution in the ASD phase, the long-term evolution is not sensitive to $P_0$. For given $\Md$ and $B_0$, there is a minimum critical $P_0$, such that the sources with $P_0$ shorter than this critical value cannot enter the ASD phase, and always remain in the SP phase. The sources with different $P_0$ evolve converging to the same period throughout the long-term ASD phase (see fig. 3 in \citet{ErtanE2009}). During the ASD phase, $\Pdot$ is not sensitive to $\Mdotin$ evolution which  depends on $\Md$ (see below). Sources with greater $\Md$ evolve with greater luminosities in the ASD phase, but the rotational evolution, which is not sensitive to $\Md$, is determined mainly by $B_0$. Large differences in $\Md$ values produce only a small scatter in the achieved current properties of the sources after the ASD/SP transition. Unlike $\Md$, small changes in $B_0$ produce rather different evolutionary paths in the ASD phase leading to a large scatter in the rotational properties after the ASD/SP transition. Figs. 1 and 2 in \citet{Ertan2014} compare the effects of $B_0$ and $\Md$ on the long-term evolutions. In this work, Figs.~\ref{fig:Lx} and \ref{fig:P_Pdot_07s} also show the sensitivity of the long-term evolution to $B_0$.
		
	In the ASD phase when the source is not very close to rotational equilibrium, the dominant torque $N_\m{D}$ is proportional to $B_0/P^2$ and independent of $\Mdotin$. Since $N_\m{TOT} = I \dot{\Omega} \propto \Pdot/P^2$, where $I$ is the moment of inertia of the neutron star, $\Pdot$ depends only on $B_0$ which we take constant during the long-term evolution. Simulations indicate that this phase terminates at an age smaller than $\sim 5 \times 10^5$~yr. The age of a source that has been evolving in the ASD phase starting from the initial phase of evolution is equal to $P/\Pdot$, comparable to characteristic age, $\tc = P/2 \Pdot$. In the SP phase following an ASD phase, $\Pdot$ is proportional to $\Mdotin$ which decreases rapidly in this late phase of evolution. Consequently, the maximum $\Pdot$ achieved in the ASD phase also decreases sharply after termination of this phase. This causes $P$ to remain almost constant after the ASD/SP transition. This maximum $P$ achieved at the end of the ASD phase depends sensitively on $B_0$, and does not change significantly with $\Md$. That is, we can estimate $B_0$ from the measured $P$ for the sources currently in the SP phase.
		
\section{Evolution of RRATs with fallback discs}
\label{sec:evol}

	For systems starting the evolution in the ASD phase, the transition to the SP phase occurs at an age smaller than about $5 \times 10^5$~yr. The age of a source is close to its characteristic age when it is in a long-lasting ASD phase. After the transition to the SP phase, $\Pdot$ decreases rapidly due to sharp decrease in $\Mdotin$. The increase in $P$ during the ASD phase stops after the ASD/SP transition, $P$ remains almost constant in the SP phase \citep[see][for details]{Ertan2014}. It is crucial for our analysis that this maximum $P$ reached at the end of the ASD phase depends basically on $B_0$, and has a weak dependence on $\Md$. This allows us to estimate $B_0$ for the RRATs that have periods sufficiently greater than their $P_0$ and $\tc \gtrsim 5 \times 10^5$~yr.
	
	These sources with $\tc \gtrsim 5 \times 10^5$~yr should be evolving in the SP phase at present. This conclusion is independent of the details of the evolution in the past. The sources with $\tc < 5 \times 10^5$~yr could be in the SP or ASD phase depending on the initial conditions. For a given source, if $\tc > 5 \times 10^5$~yr, there are two basic possibilities for the history of the source: (1) a long lasting ASD phase followed by a SP phase, (2) an evolution permanently in the SP phase. From the earlier applications of the model, we estimate that most of RRATs follow the evolutionary phases described in case (1). In this work, we investigate these solutions, and try to estimate the range of $B_0$ allowed for RRATs. For case (2), a large fraction of the sources evolve towards the properties of normal radio pulsars, while the remaining fraction enters the ASD phase at a later time of evolution, and subsequently evolves as in case (1).
	
	To estimate the RRATs evolving in the SP phase, the $\tc > 5 \times 10^5$~yr condition is reliable only for the sources with $P$ sufficiently greater than $P_0$. In the first part of our analysis, we select the sources with $P > 0.7$~s ($30$ out of $34$). It is remarkable that $\tc \gtrsim 5 \times 10^5$~yr for all these sources (except J1819). For the neutron stars starting the evolution in the ASD phase, $P_0$ does not affect the long-term evolution significantly. The periods of the sources with different $P_0$ values converge to the same level at the end of the ASD phase \citep[see e.g.][for details]{ErtanE2009}. For the model calculations, we set $P_0 = 300$~ms, the estimated average value for the young pulsars \citep{Faucher2006}. 
		
\begin{figure}
\centering
	\includegraphics[width=\columnwidth]{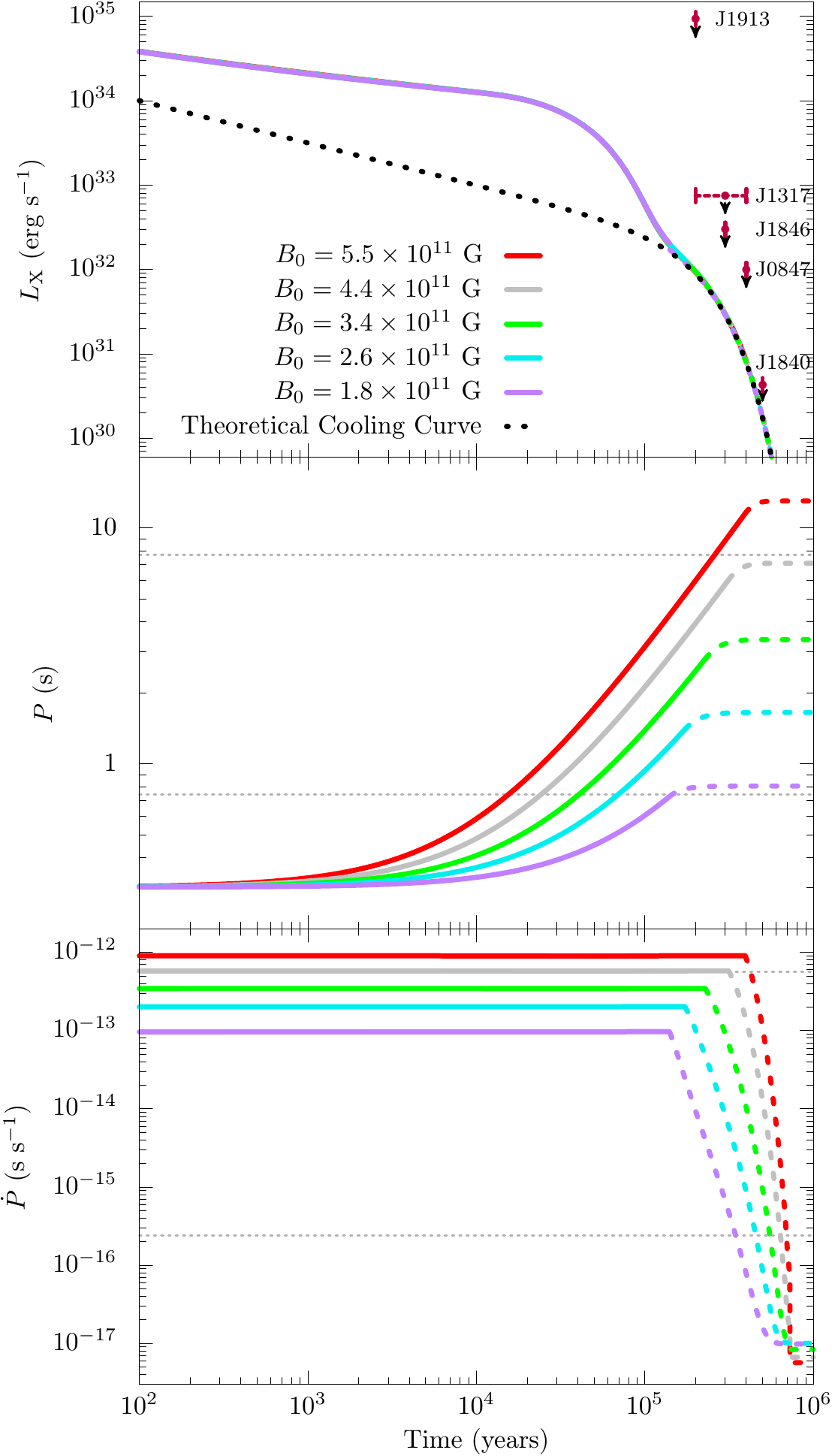}
    \caption{ The model curves produced by changing $B_0$ only (given in the top panel) for RRATs with $P > 0.7$~s. For all the curves, $\alpha = 0.045$, $\Tp = 67$~K, $C = 7 \times 10^{-4}$, $P_0 = 0.3$~s and $\Md = 3.16 \times 10^{-6}~\Msun$ (see the text for the details). The dotted curve in the top panel shows the theoretical cooling curve \citep{Page2009}. The ASD and SP phases are shown by the solid and dashed curves respectively. The X-ray upper limits of 5 RRATs and their approximate ages (see Table~\ref{tab:PPdotLx}) obtained from the model are shown with error bars in the top panel. 
    The thin horizontal lines in the middle and the bottom panels show the observed $P$ and $\Pdot$ ranges for RRATs with $P > 0.7$~s respectively.  }
    \label{fig:Lx}
\end{figure}

	Since the period achieved at the end of the ASD phase depends mainly on $B_0$ \citep[see][figs.~1 and 2]{Ertan2014}, for all model calculations, we use the same initial conditions ($\Md = 3.16 \times 10^{-6}~\Msun$ and $P_0 = 300$~ms) and the same set of main disc parameters ($\alpha = 0.045$, $\Tp = 67$~K, and $C = 7 \times 10^{-4}$), which are similar to those employed in the earlier work \citep{Ertan2006, Ertan2007, Caliskan2013, Ertan2014, Benli2016, Benli2017, Benli2018, BenliCCO2018, Gencali2018, Ozcan2020}. In this first step of our analysis, we obtain different model curves by changing the initial condition $B_0$ only. Illustrative model curves tracing the $P$ and $\Pdot$ ranges of RRATs with $P > 0.7$~s are given in Fig.~\ref{fig:Lx}. The model curves show how $P$, $\Pdot$, and $\Lx$ evolution change depending on $B_0$. The solid and dashed parts of the curves correspond to the ASD and SP phases respectively. There is a single $\Lx$ curve (top panel), since we use the same $\Md$ for all model sources. In the top panel, the dotted curve is the theoretical cooling curve for the neutron stars with conventional dipole fields. In Fig.~\ref{fig:Lx}, it is seen that all these sources are evolving in the SP phase as we have initially estimated from their characteristic ages. They are powered by the cooling luminosity at ages $(\sim 2$~--~$6) \times 10^5$~yr. 
	
\begin{figure}
\centering
	\includegraphics[width=\columnwidth]{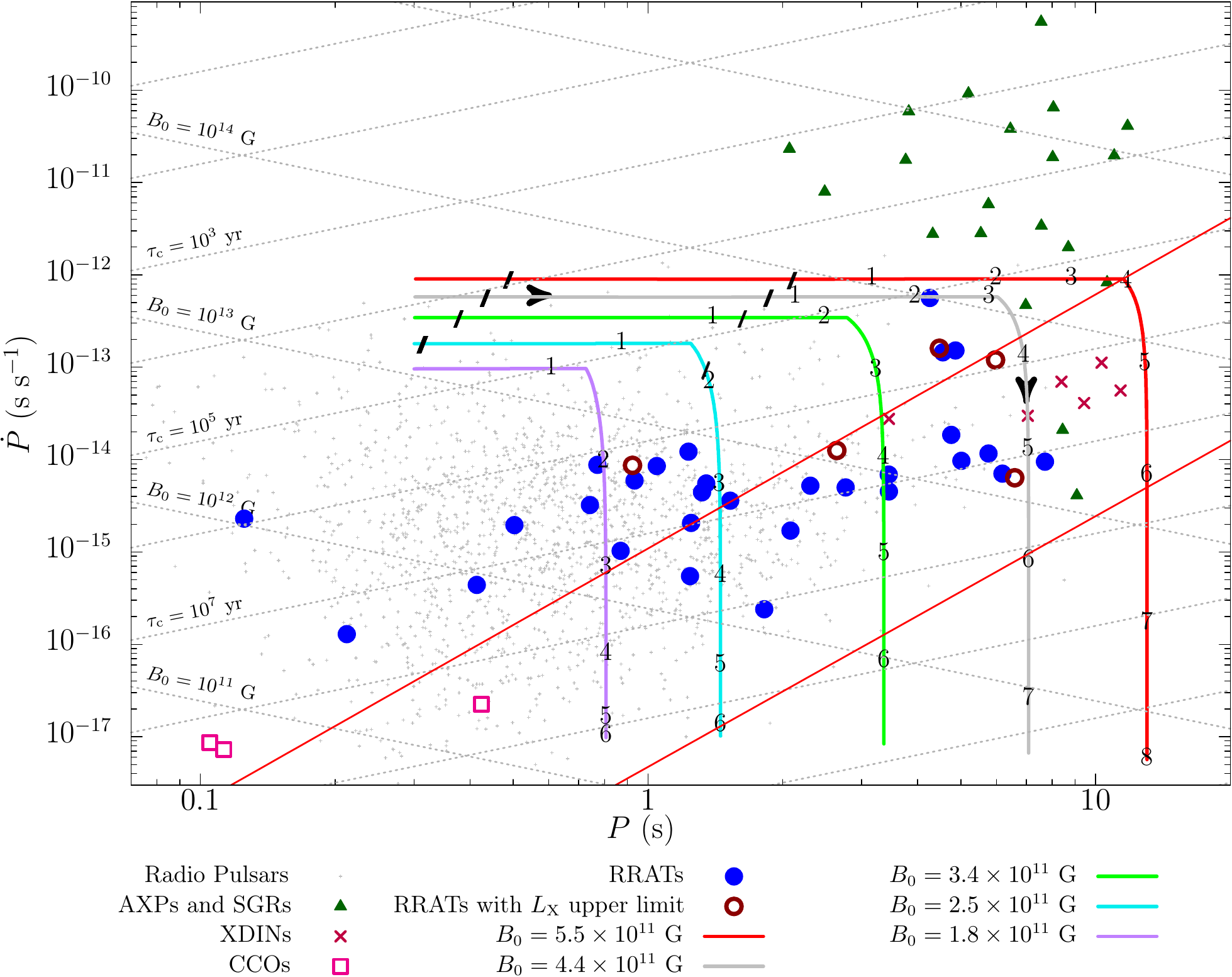}
    \caption[The LOF caption]{Evolutions in the $P$~--~$\Pdot$ diagram for the same model sources given in Fig.~\ref{fig:Lx}. Filled circles show RRATs. Unfilled circles indicate the RRATs with $\Lx$ upper limits. Red solid lines are the upper and lower boundaries of the death valley \citep{Chen1993}. Constant $\tc$ and $B_0$ calculated from the dipole-torque formula are also given for comparison with those estimated in our model (given in the figure). The numbers on the model curves show the ages of the sources in units of $10^5$~yr. The ticks on the curves correspond to the periods at which the sources cross the upper and lower borders of the death valley in our model. It is seen that the sources with $B_0 \gtrsim 3 \times 10^{11}$~G find themselves below the pulsar death line at the end of the ASD phase (horizontal branches of the curves). The constant--$B_0$ lines show the actual $B_0$ of the sources only at the end points of the model curves. Ordinary radio pulsars, AXP/SGRs, XDINs, RRATs, and CCOs are also plotted the data is taken from the ATNF Pulsar Catalogue \citep[version 1.63,][]{Manchester2005} \footnotemark. }
    \label{fig:P_Pdot_07s}
\end{figure}

	In Fig.~\ref{fig:P_Pdot_07s}, we plot the model curves for long-$P$ RRATs in the \PPdot~diagram using the same color code as in Fig.~\ref{fig:Lx}. For comparison, we kept the constant $\tc$ and $B_0$ lines estimated from the dipole torque formula. The horizontal and vertical branches of the curves in Fig.~\ref{fig:P_Pdot_07s} correspond to the ASD and SP phases respectively. The numbers seen on the model curves show the ages of the sources in the units of $10^5$~yr. It is seen in Figs.~\ref{fig:Lx} and \ref{fig:P_Pdot_07s} that the sources with stronger fields evolve to longer periods. It is remarkable that the $B_0$ values between $\sim 2 \times 10^{11}$~G and $\sim 6 \times 10^{11}$~G produce the model curves tracing the entire $P$ and $\Pdot$ ranges of all RRATs with $P > 0.7$~s. In Fig.~\ref{fig:P_Pdot_07s}, constant--$B_0$ lines show the actual $B_0$ of the sources only at the end points of the model curves where the disc torque becomes negligible compared to the dipole torque. There are $\Lx$ upper limits estimated for 5 of these RRATs. These upper limits, together with measured $P$ and $\Pdot$ of the sources are listed in Table~\ref{tab:PPdotLx}. From the model curves given in Fig.~\ref{fig:P_Pdot_07s}, the ages of these sources can be estimated. It is seen in Fig.~\ref{fig:Lx} that the cooling luminosities corresponding to these ages are in agreement with the estimated $\Lx$ upper limits.
	
\begin{figure}
\centering
	\includegraphics[width=\columnwidth]{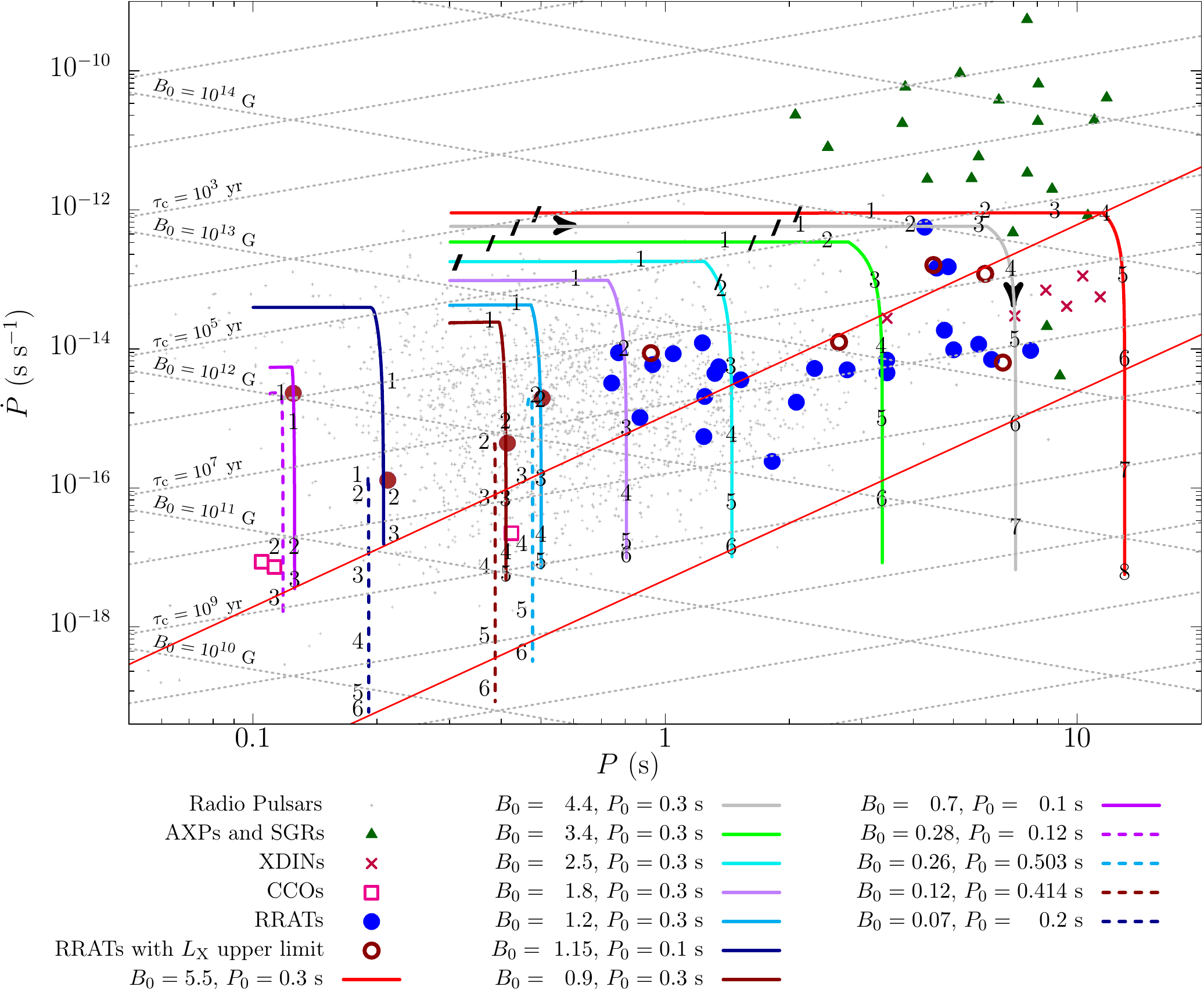}
    \caption{The same as Fig.~\ref{fig:P_Pdot_07s}, but the model curves for the sources with $P  <  0.7$~s are also added (brown filled circles). The dashed and solid curves show the evolutionary paths crossing the current source properties in the ASD and SP phases respectively. The $P_0$ and $B_0$ values in the unit of $10^{11}$~G are given in the figure.}
    \label{fig:P_Pdot}
\end{figure}

	Among the RRATs with measured $\Pdot$, the other sources have periods in the $0.1$~--~$0.5$~s range.
	For these sources, we cannot estimate the evolutionary phase from the $\tc$, since their $P$ could be close to $P_0$. 
	Nevertheless, for a particular source, we could estimate two different $B_0$ values for each of the two possible phases of evolution, ASD and SP, at present. This analysis is important in that it is likely to give the minimum possible dipole fields for the RRAT population in our model. For these sources, we leave $P_0$ free in addition to $B_0$, and do not change $\Md$ and the main disc parameters used for the $P > 0.7$~s sources. In Fig.~\ref{fig:P_Pdot}, the model curves obtained for all RRATs are seen together. The model fits indicate that the sources with $P < 0.7$~s could have $B_0$ ranging from $\sim 7 \times 10^9$~G to $\sim 10^{11}$~G. For a given source, the minimum possible $B_0$ is estimated assuming that the source in the ASD phase at present. We estimated these minimum possible $B_0$ values in our earlier work \citep[see][ fig.~2]{Gencali2018}.
	
\begin{figure}
\centering
	\includegraphics[width=\columnwidth]{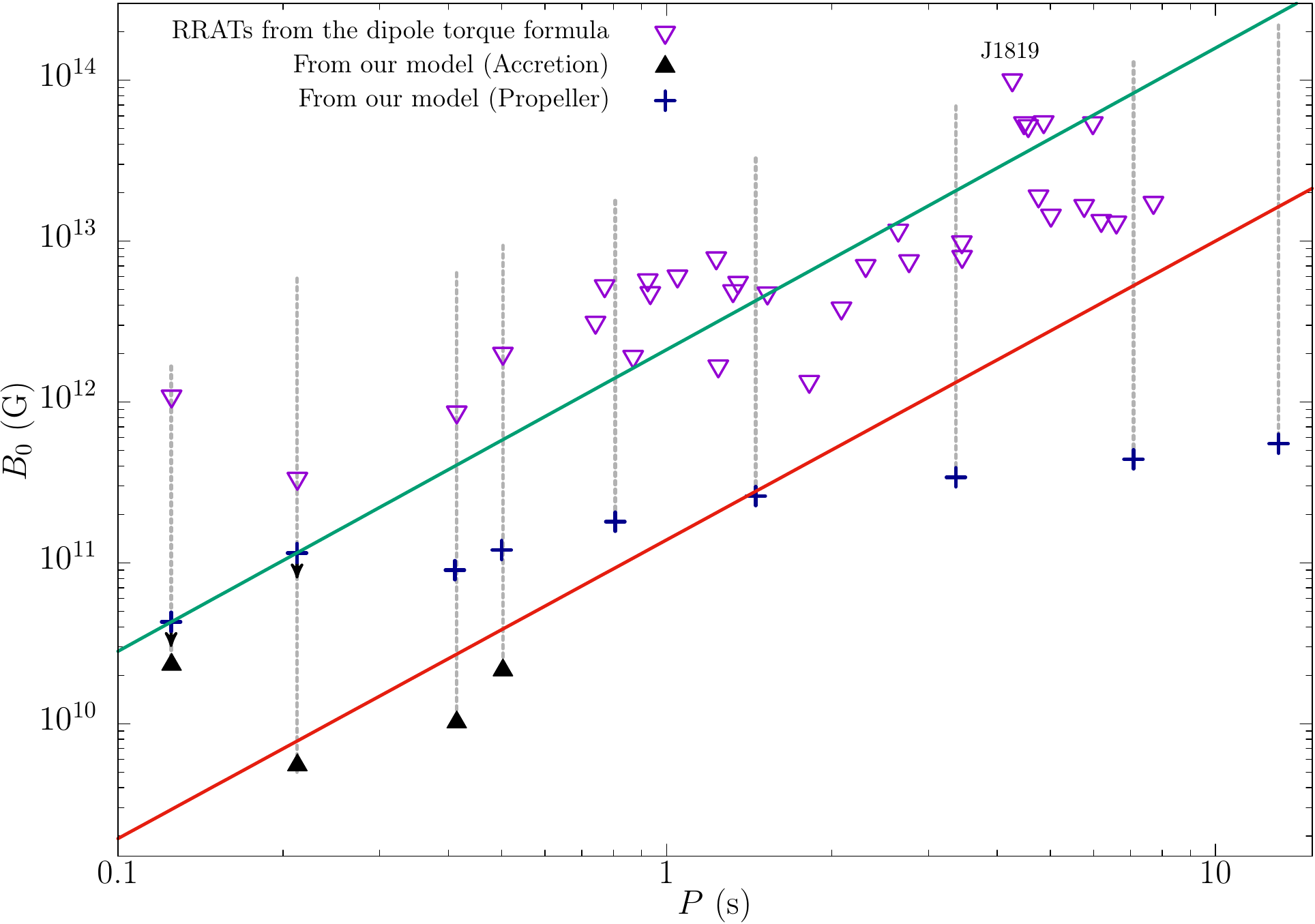}
    \caption{RRATs in the $B_0$~--~$P$ diagram. Plus signs and filled triangles show the $B_0$ values and final periods for the evolutions seen in Fig.~\ref{fig:P_Pdot}. 
The solid lines show the upper and lower borders of the death valley \citep{Chen1993}. Empty triangles indicate the $B_0$ values inferred from the dipole-torque formula using measured $P$ and $\Pdot$ values (see Table~\ref{tab:PPdotLx}). Vertical dashed lines show the ranges of $B_0$ deduced from the dipole torque formula during the long-term evolution. For the four sources with $P < 0.7$~s, plus signs and filled triangles show the locations in the case that the sources are currently in the SP phase and the ASD phase respectively. For $P > 0.7$~s, there is only one source (J1819) in the the ASD phase, while the others are in the SP phase.}
    \label{fig:B0_P}
\end{figure}

	The $B_0$ values estimated in our model are also given in the \B0-P diagram in Fig.~\ref{fig:B0_P}. The solid lines show the upper and lower boundary of the pulsar death valley \citep{Chen1993}. The $B_0$ values inferred from the dipole torque formula are also plotted for comparison (empty triangles). Filled triangles show $B_0$ values of the sources with $P < 0.7$~s in the case that they are currently in the ASD phase. The solutions producing the source properties in the SP phase can be obtained with greater $B_0$ depending on $P_0$. For the two sources with smallest $P$, the upper $B_0$ values given in Figs.~\ref{fig:P_Pdot} and \ref{fig:B0_P} correspond to the maximum $B_0$ leaving the sources inside the death valley. We estimate that the death point of the RRAT with minimum $P$ is close to the upper border of the death valley (see Section~\ref{sec:conc}). It is seen in Fig.~\ref{fig:B0_P} that all these sources could be located inside the death valley or below the pulsar death line with the $B_0$ values estimated in our model.

\footnotetext{\url{https://www.atnf.csiro.au/research/pulsar/psrcat/}}
	
	The estimated ages and $\Lx$ values in our model together with the observed rotational properties are given in Table~\ref{tab:PPdotLx}. As seen in Fig.~\ref{fig:Lx}, most of the RRATs are powered by the intrinsic cooling of the neutron star. We should note that the cooling curve depends on some unknown details of the neutron star properties, such as the equation of state and mass \citep[see e.g.][for a recent detailed discussion]{Potekhin2020}. In our numerical calculations, we have employed the cooling curve obtained by \citet{Page2009}. Any corrections to the cooling curve could change the $\Lx$ values estimated in our illustrative model, without a significant change in the $B_0$ values. 
	
	The ages of most RRATs estimated in our model are one to two orders of magnitude smaller than their characteristic ages. This indicates that RRATs could have cooling luminosities in soft X-rays that are much higher than those corresponding to their characteristic ages and comparable to those of XDINs. This puts a significant fraction of RRATs, especially those within $2$~kpc, into the list of detectable sources for the $4$-year extended Roentgen Survey with an Imaging Telescope Array (eROSITA) all-sky survey (eRASS) which started on December 13, 2019  \citep{Merloni2012, Pires2017}. All known XDINs are within a distance of $500$~pc from the Sun, while most of the RRATs are at larger distances (see Table~\ref{tab:PPdotLx}). Nevertheless, the eRASS sensitivity will be about $25$ times greater than the ROSAT all-sky survey sensitivity in the $0.2$~--~$2.3$~keV soft X-ray range \citep{Predehl2020}. Through Monte Carlo simulations, \citet{Pires2017} estimated that, on average, about $90$ thermally emitting new isolated neutron stars could be detected by the eRASS.
				
\section{Discussion \& Conclusions}
\label{sec:conc}
	
	We have estimated the $B_0$ range for RRATs with measured $\Pdot$ in the fallback disc model. Comparing our results with those obtained earlier for the other young neutron star populations, we see that the $B_0$ range of RRATs ($\sim 7 \times 10^9$~--~$\sim 6 \times 10^{11}$~G) could fill the gap between the $B_0$ ranges of CCOs and XDINs ($\sim 4 \times 10^9$~--~$\sim 3 \times 10^{11}$~G) estimated in the same model \citep{Ertan2014, BenliCCO2018}.
	
	In the dipole torque model, almost all RRATs are located above the pulsar death line, and about half of them are even above the upper border of the pulsar death valley in the \B0-P diagram. With the $B_0$ values estimated in our model, we find that all these sources could be either inside the death valley or below the death line. Given that RRATs do not show strong continuous radio pulses, we estimate that they are close to or below their individual death points. Then, what could be the reason for their repetitive, short radio bursts? We tentatively propose that these bursts could be ignited through interaction of the inner disc with the dipole field of the star at the inner boundary of the disc. The flux of the open field lines through the magnetic poles could be occasionally enhanced during the continual opening and reconnection of the lines passing through the inner disc boundary \citep{Lovelace1995, Lovelace1999}.
	
	The enhancement of the open field line flux was proposed earlier by \citet{Parfrey2016} in a different context related to the effect of this enhancement on the torques acting on the millisecond pulsars. For the neutron stars with long $P$, like RRATs, contribution of the field-line opening to the spin-down torque is negligible compared to the disc torque. Nevertheless, we estimate that this mechanism could transiently build up the condition required for the strong radio emission near the poles. The field lines, that are closed inside the light cylinder radius of a neutron star rotating in vacuum, open up from $\rlc$ to $\sim \rin$ when there is a disc around the star. Outside the inner boundary the lines are estimated to be decoupled from the disc, while the lines passing through the boundary interact with the disc. This interaction inflates and opens up the closed lines on a short interaction time scale. The reconnections of the lines are also estimated to occur on a similar time-scale \citep{Lovelace1999, Ustyugova2006}. As a result of these cycles, the field density and the flux of the open field lines through the poles are likely to show continual variations, which might trigger the observed brief radio bursts of RRATs when the appropriate condition for the burst is built up for short time intervals.
	
	In this picture, the disc torque should be sufficiently strong to ignite radio bursts when the star is actually below its death point. For instance, XDINs have relatively long periods and are estimated to be in the strong-propeller phase with fields placing them well below the pulsar death line in the fallback disc model \citep{Ertan2014}. This might indicate that their disc torques are not strong enough to produce radio bursts. In other words, we estimate that the RRAT behavior is not likely to be observed beyond a certain period. Below this period, there could be a $P$ range along which both XDINs and RRATs could show radio bursts depending on their locations of the death points. The fact that 5 of XDINs have periods greater than the maximum period ($\sim 7$~s) of RRATs seems to support this idea. For the evolutions that end up inside the death valley, a given source could show normal radio pulses or RRAT behavior depending on the position of its death point inside the valley. For instance, the model source with $B_0 = 1.8 \times 10^{11}$~G and $P_0 = 0.3$~s (Fig.~\ref{fig:P_Pdot}), is estimated to be inside the valley after the inactivation of the disc (at the end point of the model curve). If its death point is below the end point of the curve, the source could be a radio pulsar during and after the SP phase. If the death point is above the end point, the source could be observed as an RRAT in the SP phase. In this picture, the effect of the disc is to boost the source transiently above its death point inside the valley. This affect is strongest at the beginning of the SP phase when $\rin \simeq \rco$, and becomes negligible towards the end points of the evolutionary curves (Fig.~\ref{fig:P_Pdot}). The radio bursts are likely to terminate when the flux enhancements become insufficient to trigger radio bursts. This is likely to occur when a given source is apparently either inside the valley or possibly, for low-period sources, beyond the upper border of the valley. Investigation of the conditions that could trigger radio bursts, which is beyond the scope of this work, could be studied through numerical simulations of the disc-field interaction.
		
	Considering that RRATs have relatively high birth rate among the young neutron star populations \citep{Popov2006}, our results imply that a large fraction of the neutron stars are born with $B_0$ in the $(1$~--~$10) \times 10^{11}$~G range. Among these sources, those with relatively high $B_0$ tend to evolve to longer periods leaving them well below the pulsar death line at the end of the ASD phase. These sources are likely to have XDIN properties, and unlikely to show the RRAT behavior. The sources with relatively weak fields, could enter the radio pulsar or RRAT phase depending on the positions of their death points inside the death valley. If the disc-field interaction is indeed required for the RRAT bursts, then the duration of the RRAT phase could be limited by the active lifetime of the disc which we estimate to be less than $10^6$~yr. For all RRATs plotted in Fig.~\ref{fig:P_Pdot}, the ages are in the range of $(\sim 2$~--~$6) \times 10^5$~yr in our model, while the characteristic ages are mostly greater than $10^6$~yr. At ages $\gtrsim 10^6$~yr, RRATs are not likely to be detected in X-rays since the cooling luminosities remain below the X-ray detection limits. With the ages estimated in our model (Fig.~\ref{fig:P_Pdot}), some of the RRATs could be detected with $\Lx = \Lcool \sim (10^{31}$~--~$10^{32})$~\ergpers. For the five RRATs, the estimated $\Lx$ upper limits (Table~\ref{tab:PPdotLx}), are consistent with the cooling luminosities corresponding to their ages estimated in our model (Figs.~\ref{fig:Lx} and \ref{fig:P_Pdot_07s}). 
	
	For the evolutionary curves seen in Fig.~\ref{fig:P_Pdot_07s}, all RRATs with $P > 0.7$~s are expected to be in the SP phase with one exception, namely J1819. This is the only source with known $\Lx$, and recently we estimated that the source is in the ASD phase, and close to the ASD/SP transition at present \citep{Gencali2018}. Continuous radio pulsations are estimated to be switched off by mass-flow on to the neutron star. Nevertheless, it is not clear whether very short radio bursts could be emitted in the accretion phase. For the accreting sources close to the ASD/SP transition, the short radio bursts might be emitted when the accretion is occasionally hindered. We are not sure about the current phases of the RRATs with smallest periods. Except these one or two sources that are close to the ASD/SP transition, we do not find any RRAT in the ASD phase. Finally, we note that the 3 CCOs given in Figs.~\ref{fig:P_Pdot_07s} and \ref{fig:P_Pdot} are estimated to be in the accretion phase at present with $B_0 \sim (1$~--~$5) \times 10^9$~G in the fallback disc model \citep{BenliCCO2018}. The simulations show that their periods will not change significantly until the ASD/SP transition. This means that these sources are likely to evolve into the RRAT phase after or close to the termination of the ASD phase. We will investigate the evolutionary links between these isolated neutron star populations in an independent work.
	
\section{Summary of Results}
\label{sec:SumofResults}	
	
	We have shown that the evolutionary avenues of the model sources depend mainly on the dipole field strength, $B_0$, which we have estimated to have a continuous distribution from a few $10^9$~G to about $10^{13}$~G. As seen in Fig.~\ref{fig:P_Pdot}, the sources with lowest $B_0$ values are born as either CCOs accreting matter from the disc or RRATs in the SP phase. After termination of the ASD phase, CCOs could become radio pulsar or RRAT depending on its $B_0$, current $P$ and the position of death point inside the death valley. We estimate that a large fraction of the sources have $B_0$ in the rage of ($\sim 10^{11}$~--~$10^{12}$)~G. These sources evolve to longer periods in the ASD phase, during which some of them could be identified as AXPs due to their apparently high $B_0$ inferred from the dipole torque formula (see Fig.~\ref{fig:P_Pdot_07s}). When the accretion is switched off, a large fraction of these neutron stars could enter the SP phase as RRATs. Among these systems, those with relatively short periods ($\lesssim 1$~s) could become either radio pulsar or RRAT depending on the positions of their death points inside the death valley. The stronger dipole fields lead the systems to longer periods during the ASD phase. Sources reaching the ASD/SP transition with periods longer than a critical level are likely to become XDIN immediately after the transition to the SP phase. It is seen in Fig.~\ref{fig:P_Pdot_07s} that the periods of RRATs and XDINs overlap in roughly in the  $\sim 3$~--~$8$~s range. It is possible that the XDINs in this range could actually be RRATs with unseen radio bursts due to narrow beaming and viewing geometry, or alternatively their RRAT phases terminated at an earlier time, and now they are evolving in the XDIN phase emitting no radio bursts.
	
	In the same model, it is found that a large fraction of HBRPs and persistent AXP/SGRs have $B_0$ in the ($10^{12}$~--~$10^{13}$)~G range \citep{Benli2015, Benli2016, Benli2017, Benli2018}, greater than the $B_0$ values estimated for CCOs, RRATs, and XDINs \citep{Ertan2014, BenliCCO2018}. Most of these HBRPs with relatively strong fields evolve from the SP phase to the ASD phase achieving the AXP/SGR properties. There are also HBRPs evolving always in the SP phase apparently converging to the properties of normal radio pulsars \citep{Benli2017, Benli2018}. A fraction of these neutron stars could reach their death points while the disc torques are still efficient. They become RRATs during their late phase of evolution. The remaining fraction of these HBRPs always evolve as radio pulsars slowing down initially with the disc torques in the SP phase, and eventually with dipole torques until they reach their critical periods switching off the radio pulses. Note that, there is no evolution from the RRAT to the radio pulsar phase. At the end of the RRAT phase, sources find themselves below their death points in the \B0-P diagram.
	
	We estimate that most of the AXP/SGRs are currently evolving in the ASD phase. In this population, sources with $B_0 > 10^{12}$~G could achieve periods longer than XDIN periods at the end of their ASD phases due to their relatively strong dipole fields \citep{Benli2016}. These sources are not likely to show radio pulses or bursts after the ASD phase due to their long periods placing them well below the pulsar death line. On the other hand, most of the transient AXPs, which seem to have a birth rate greater than that of persistent AXP/SGRs \citep{KeaneK2008}, could have relatively weak fields ($B_0 \lesssim 10^{12}$~G), and evolve first into the RRAT phase and later become XDIN, or possibly evolve directly into the XDIN phase if they reach sufficiently long periods at the end of the ASD phase.
	
	To sum up, in our model, RRATs seem to have evolutionary connections with all the other isolated neutron star populations. Different systems show RRAT behavior along a certain late phase of their observable long-term evolution. These connections, complicating the total birth-rate estimation of isolated neutron stars, imply that the total birth rate could be a small fraction of  the sum of the birth rates estimated for each population.

\begin{table*}
\begin{center}
\caption{ RRATs with the quantities estimated in the model (Age and $\Lx$) together with the observed properties ($P$, $\Pdot$, $d$, $L_\m{X,upper}$, $\dot{E}$ and $\tau_\m{c}$) \citep[ATNF Pulsar Catalogue, version 1.63,][]{Manchester2005, Keane2011,Bates2012}. $B_\m{db}$ shows the dipole field strengths at the pole of the star inferred with dipole braking assumption (see Fig.~\ref{fig:P_Pdot} for $B_0$ values estimated in our model). $L_\m{X,upper}$ shows the X-ray upper limits in the $0.3~-~5$~keV energy band for J1913+1330 and J1317--5759 \citep{Rea2008}, and in the $0.3~-~8$~keV energy band for J1846--0257 and J0847--4316 \citep{Kaplan2009}. $L_\m{X,upper}$ for J1840--1419 is calculated from blackbody temperature upper limit \citep{Keane2013}. Unlike other long-$P$ RRATs, J1819 is estimated to be in the ASD phase and its $\Lx$ value is obtained from detailed model fits in earlier work \citep[for details see][]{Gencali2018}. }
\label{tab:PPdotLx}
\begin{tabular}{ l c c c c c c c c c } 
 \hline
Source Name   &    $P$    &           $\Pdot$         &    $d$    &    $L_\m{X,upper}$        &       $\dot{E}$           &   $\tau_\m{c}$  &    $B_\m{db}$     &      Age        &             $L_\m{X}$        \\ 
              &     (s)   &  ($10^{-15}$~s~s$^{-1}$)  &   (kpc)   &  ($10^{32}$~erg~s$^{-1}$) &  ($10^{32}$~erg~s$^{-1}$) &   ($10^5$~yr)   &   ($10^{12}$~G)   &   ($10^5$~yr)   &   ($10^{30}$~erg~s$^{-1}$)   \\[1ex] 
\hline  
J1554--5209    &  $0.13$  &           $2.29$          &  $3.08$	  &        $-$                &          $460$            &     $8.65$      &     $0.54$        &       $1$	      & 	      $240$         \\
J1647--3607    &  $0.21$  &           $0.13$          &  $15.94$  &        $-$                &          $5.30$           &     $261$      &     $0.17$        &       $2$	      & 	      $90$          \\
J1848--1243    &  $0.41$  &           $0.44$          &  $3.15$	  &        $-$                &          $2.40$           &     $149$       &     $0.43$        &       $2$	      & 	      $90$          \\
J1623--0841    &  $0.50$  &           $1.96$          &  $4.64$	  &        $-$                &          $6.10$           &     $40.8$      &     $1$           &       $2$	      & 	      $90$          \\
J1909+0641     &  $0.74$  &           $3.22$          &  $1.33$	  &        $-$                &          $3.10$           &     $36.5$      &     $1.56$        &       $2~-~3$	  & 	      $(30~-~90)$   \\
J1826--1419    &  $0.77$  &           $8.78$          &  $3.24$	  &        $-$                &          $7.60$           &     $13.9$      &     $2.63$        &       $2$	      & 	      $90$          \\ 
J2325--0530    &  $0.87$  &           $1.03$          &  $1.49$	  &        $-$                &          $0.62$           &     $134$       &     $0.96$        &       $3$	      & 	      $30$          \\ 
J1913+1330     &  $0.92$  &           $8.68$          &  $6.18$	  &       $<~940$             &          $4.40$           &     $16.9$      &     $2.86$        &       $2$  	  & 	      $90$          \\ 
J1839--0141    &  $0.93$  &           $5.94$          &  $6.07$	  &        $-$                &          $2.90$           &     $24.9$      &     $2.38$        &       $2~-~3$   & 	      $(30~-~90)$   \\ 
J1513--5946    &  $1.05$  &           $8.53$          &  $3.76$	  &        $-$                &          $2.90$           &     $19.4$      &     $3.02$        &       $2~-~3$   & 	      $(30~-~90)$   \\ 
J1048--5838    &  $1.23$  &           $12.2$          &  $1.01$	  &        $-$                &          $2.60$           &     $16$        &     $3.92$        &       $2~-~3$   & 	      $(30~-~90)$   \\ 
J0628+0909     &  $1.24$  &           $0.55$          &  $1.77$	  &        $-$                &          $0.11$           &     $359$       &     $0.84$        &       $4$	      & 	      $8$           \\ 
J0139+3336     &  $1.25$  &           $2.06$          &  $1.47$	  &        $-$                &          $0.42$           &     $95.8$      &     $1.62$        &       $3$       & 	      $30$          \\ 
J1754--3014    &  $1.32$  &           $4.43$          &  $2.52$	  &        $-$                &          $0.76$           &     $47.2$      &     $2.45$        &       $3~-~4$	  & 	      $(8~-~30)$    \\ 
J0201+7005     &  $1.35$  &           $5.51$          &  $1.17$	  &        $-$                &          $0.89$           &     $38.8$      &     $2.76$        &       $3$	      & 	      $30$          \\ 
J0912--3851    &  $1.53$  &           $3.59$          &  $0.52$	  &        $-$                &          $0.40$           &     $67.4$      &     $2.37$        &       $3$	      & 	      $30$          \\ 
J1739--2521    &  $1.82$  &           $0.24$          &  $4.41$	  &        $-$                &          $0.02$           &     $1200$      &     $0.67$        &       $4~-~5$	  & 	      $(2~-~8)$     \\ 
J1919+1745     &  $2.08$  &           $1.71$          &  $4.09$	  &        $-$                &          $0.08$           &     $193$       &     $1.91$        &       $4$       & 	      $8$           \\ 
J2105+6223     &  $2.31$  &           $5.22$          &  $2.72$	  &        $-$                &          $0.17$           &     $70$        &     $3.51$        &       $3~-~4$   & 	      $(8~-~30)$    \\ 
J1317--5759    &  $2.64$  &           $12.6$          &  $3.77$	  &       $<~7.5$             &          $0.27$           &     $33.3$      &     $5.83$        &       $3~-~4$	  & 	      $(8~-~30)$    \\ 
J1807--2557    &  $2.76$  &           $4.99$          &  $18.56$  &        $-$                &          $0.09$           &     $87.7$      &     $3.76$        &       $4$	      & 	      $8$           \\ 
J1538+2345     &  $3.45$  &           $6.89$          &  $1.31$	  &        $-$                &          $0.07$           &     $79.3$      &     $4.93$        &       $4$	      & 	      $8$           \\ 
J1854--1557    &  $3.45$  &           $4.52$          &  $10.90$  &        $-$                &          $0.04$           &     $121$       &     $4$           &       $4~-~5$	  & 	      $(2~-~8)$     \\ 
J1819--1458    &  $4.26$  &           $563$           &  $3.30$	  &        $-$                &          $2.90$           &     $1.2$       &     $49.6$        &       $2$	      & 	      $4000$          \\ 
J1846--0257    &  $4.48$  &           $161$           &  $4.00$	  &        $<~3$              &          $0.71$           &     $4.42$      &     $27.1$        &       $3$	      & 	      $30$          \\ 
J1854+0306     &  $4.56$  &           $145$           &  $4.50$	  &        $-$                &          $0.61$           &     $4.98$      &     $26$          &       $3$	      & 	      $30$          \\ 
J1444--6026    &  $4.76$  &           $18.5$          &  $5.82$	  &        $-$                &          $0.07$           &     $40.7$      &     $9.51$        &       $4$	      & 	      $8$           \\ 
J0736--6304    &  $4.86$  &           $152$           &  $0.90$	  &        $-$                &          $0.52$           &     $5.07$      &     $27.5$        &       $3~-~4$	  & 	      $(8~-~30)$    \\ 
J2033+0042     &  $5.01$  &           $9.70$          &  $2.93$	  &        $-$                &          $0.03$           &     $81.9$      &     $7.06$        &       $4~-~5$	  & 	      $(2~-~8)$     \\ 
J1707--4417    &  $5.76$  &           $11.7$          &  $14.05$  &        $-$                &          $0.02$           &     $78.4$      &     $8.29$        &       $5$	      & 	      $2$           \\ 
J0847--4316    &  $5.98$  &           $120$           &  $6.55$	  &        $<~1$              &          $0.22$           &     $7.9$       &     $27.1$        &       $4$	      & 	      $8$           \\ 
J1226--3223    &  $6.19$  &           $7.05$          &  $2.17$	  &        $-$                &          $0.01$           &     $139$       &     $6.69$        &       $5$       & 	      $2$           \\ 
J1840--1419    &  $6.60$  &           $6.35$          &  $0.72$	  &        $<~0.043$          &          $0.009$          &     $165$       &     $6.55$        &       $5$	      & 	      $2$           \\ 
J1652--4406    &  $7.70$  &           $9.50$          &  $5.11$	  &        $-$                &          $0.008$          &     $129$       &     $8.66$        &       $5$	      & 	      $2$           \\ 
 \hline 
 \end{tabular}
\end{center}
\end{table*}

\section*{Acknowledgements}

	We thank the referee, Adriana Mancini Pires, for very useful comments that have considerably improved our manuscript. We acknowledge research support from Sabanc{\i} University, and from T\"{U}B\.{I}TAK (The Scientific and Technological Research Council of Turkey) through grant 117F144.

\section*{DATA AVAILABILITY}

	The data underlying this article will be shared on reasonable request to the corresponding author.



\bibliographystyle{mnras}
\bibliography{mnras} 








\bsp	
\label{lastpage}
\end{document}